\documentstyle[preprint,eqsecnum,aps]{revtex}
\tighten
\begin{document}
\draft
\preprint{gr-qc/9911069}
\title{Uniqueness theorems for static black holes in metric--affine gravity}
\author{Eloy Ay\'on--Beato$^{\P}$\thanks{%
E-mail: ayon@fis.cinvestav.mx}, Alberto Garc\'{\i}a$^{\P}$\thanks{
E-mail: aagarcia@fis.cinvestav.mx}, Alfredo Mac\'{\i}as$^{\diamond\P}$
\thanks{
E-mail: amac@xanum.uam.mx; macias@fis.cinvestav.mx}, and Hernando Quevedo$
^{\star}$\thanks{
E-mail: quevedo@nuclecu.unam.mx}}
\address{$^{\P}$~Departamento~de~F\'{\i}sica,~CINVESTAV--IPN,\\
Apartado~Postal~14--740,~C.P.~07000,~M\'{e}xico,~D.F.,~MEXICO.\\
$^{\diamond}$~Departamento~de~F\'{\i}sica,~Universidad~Aut%
\'onoma~Metropolitana--Iztapalapa\\
Apartado~Postal~55--534,~C.P.~09340,~M\'{e}xico,~D.F.,~MEXICO.\\
$^{\star}$~Instituto~de~Ciencias~Nucleares,~Universidad~Nacional~Aut%
\'onoma~de~M\'exico\\
Apartado~Postal~70--543,~C.P.~04510,~M\'{e}xico,~D.F.,~MEXICO.}
\date{\today}
\maketitle

\begin{abstract}
Using the equivalence theorem for the {\it triplet ansatz\/} sector of
metric--affine gravity (MAG) theories and the Einstein--Proca system, it is
shown that the only static black hole of the triplet sector of MAG is the
Schwarzschild solution, under the constraint $(-4\beta _{4}+k_{1}\beta
_{5}/2k_{0}+k_{2}\gamma _{4}/k_{0})/\kappa z_{4}\neq 0$ on the coupling
constants. For the special case $(-4\beta _{4}+k_{1}\beta
_{5}/2k_{0}+k_{2}\gamma _{4}/k_{0})/\kappa z_{4}=0$, it follows that the
only static non--extremal black hole is the Reissner--Nordstr\"{o}m one. The
results can be extended to exclude also the existence of soliton solutions
of the triplet sector of MAG. 
\end{abstract}



\section{Introduction}

In February 1916, only three months after having achieved the final
breakthrough in general relativity, Einstein presented \cite{schw}, on
behalf of Schwarzschild, the first exact solution of his new equations to
the Prussian Academy of Sciences. However, it took almost half a century
until the geometry of the Schwarzschild space--time was correctly
interpreted and its physical significance was fully appreciated. The vacuum
Schwarzschild solution describing the end product of gravitational collapse
contains a space--time singularity which is hidden within a black hole.

The mathematical theory of black holes has been steadily developing during
the last thirty years. One of the most intriguing outcomes is the so--called
``{\em no--hair}'' theorem, which states that a black hole in a stationary
electrovacuum space--time is uniquely characterized by its mass, angular
momentum and electric charge (see Refs. \cite{Heusler96,Heusler98,Bekens98}
for recent reviews on the subject).

Although Einstein gravity is well founded and experimentally very successful
on the macroscopic scale, it is of interest to investigate gravity theories
which generalize the restricted geometrical structure of Einstein's theory.
The theory of the quantum superstring \cite{quantum}, suggests that
non--Riemannian features are present on the scale of the Planck length.
Since the possibility of testing the predictions of such fundamental
theories is very low, due to the high energy levels where new effects may
appear, effective theories offer an alternative scenario to perform tests at
lower energy levels. It turns out that low--energy dilaton and axi--dilaton
interactions are manageable in terms of a non--Riemannian connection that
leads to new geometrical structures with particular torsion and nonmetricity
fields \cite{nehe}.

More generally, the metric--affine theory of gravity (MAG), which is
basically a gauge theory of the four--dimensional affine group, encompasses
the dilaton--axion gravity of the effective string models as a subcase \cite
{dt95,dot95}. MAG has exact solutions with novel features \cite
{PDMAG,cwMAG,electrovacMAG,TresguerresShear1} (For a review on exact
solutions in MAG see Ref. \cite{hema99} and references therein). These
solutions and a possible necessary symmetry reduction process may pave a way
to understand how the Riemann--Einstein structure emerges in gravitational
gauge theory.

In MAG, the nonmetricity, torsion and curvature are dynamical variables ---
field strengths --- which together with the coframe, the metric and the
connection potentials provide an alternative description of gravitational
physics.

Black holes with non--Riemannian geometrical structures have not been
extensively studied in the literature. Tresguerres \cite{TresguerresShear1}
and Tucker and Wang \cite{twbh} found Reissner--Nordstr\"om--like metrics,
in which the place of the electric charge is taken by the {\em dilation
charge}, (gravito--electric charge) related to the Weyl covector, the trace
of the nonmetricity, together with a vector part of the torsion, i.e., these
solutions carry a {\em dilation} charge (Weyl charge) and a {\em spin}
charge, but are devoid of any other post--Riemannian ``excitations,'' in
particular, they have no tracefree pieces of the nonmetricity. The
corresponding Lagrangian needs only a Hilbert--Einstein piece and a
segmental curvature squared. Other black hole solutions exciting more
post--Riemannian structures have been found by Vlachinsky et al. \cite
{heh961} and Obukhov et al. \cite{heh96}. However, the most general
Reissner--Nordstr\"om solution in MAG, endowed with electric and magnetic
charges, as well as gravito--electric and gravito--magnetic charges and
cosmological constant term is presented in \cite{malo}.

In this paper, by using the equivalence theorem between the {\em triplet
ansatz} sector of MAG and the Einstein--Proca systems \cite{obu,tw}, we
proof a ``{\em no--hair}'' theorem for static black holes in this sector of
MAG theories. The result implies that the only static black hole of the
triplet ansatz sector of MAG, in vacuum is the Schwarzschild solution and,
in electrovacuum is the Reissner--Nordstr\"om black hole.

In Section II we review the main aspects of MAG and its {\em triplet ansatz}
sector together with the equivalence theorem to the Einstein--Proca systems.
In Section III the black hole configurations in MAG are reviewed and
discussed. In Section VI under the assumption that it represents the
gravitational field of a black hole, we analyze the field equations for a
static configuration . We proof a ``no--hair'' theorem for the effective
Proca field derived from the equivalence theorem. In Section V we discuss
the physical significance of our results.


\section{MAG in brief and the equivalence theorem}

The {\em most general parity conserving quadratic} Lagrangian which is
expressed in terms of the $4+3+11$ irreducible pieces (see \cite{PR,hema99})
of $Q_{\alpha \beta }$, $T^{\alpha }$, $R_{\alpha }{}^{\beta }$,
respectively, reads: 
\begin{eqnarray}
V_{{\rm MAG}} &=&\frac{1}{2\kappa }\,\left[ -a_{0}\,R^{\alpha \beta }\wedge
\eta _{\alpha \beta }+T^{\alpha }\wedge {}^{\ast }\!\left(
\sum_{I=1}^{3}a_{I}\,^{(I)}T_{\alpha }\right) \right.  \nonumber \\
&+&\left. 2\left( \sum_{I=2}^{4}c_{I}\,^{(I)}Q_{\alpha \beta }\right) \wedge
\vartheta ^{\alpha }\wedge {}^{\ast }\!\,T^{\beta }+Q_{\alpha \beta }\wedge
{}^{\ast }\!\left( \sum_{I=1}^{4}b_{I}\,^{(I)}Q^{\alpha \beta }\right)
\right.  \nonumber \\
&+&b_{5}\bigg.\left( ^{(3)}Q_{\alpha \gamma }\wedge \vartheta ^{\alpha
}\right) \wedge {}^{\ast }\!\left( ^{(4)}Q^{\beta \gamma }\wedge \vartheta
_{\beta }\right) \bigg]  \label{lagr} \\
&-&\frac{1}{2\rho }\,R^{\alpha \beta }\wedge {}^{\ast }\!\left(
\sum_{I=1}^{6}w_{I}\,^{(I)}W_{\alpha \beta }+w_{7}\,\vartheta _{\alpha
}\wedge (e_{\gamma }\rfloor ^{(5)}W^{\gamma }{}_{\beta })\right.  \nonumber
\\
&+&\left. \sum_{I=1}^{5}{z}_{I}\,^{(I)}Z_{\alpha \beta }+z_{6}\,\vartheta
_{\gamma }\wedge (e_{\alpha }\rfloor ^{(2)}Z^{\gamma }{}_{\beta
})+\sum_{I=7}^{9}z_{I}\,\vartheta _{\alpha }\wedge (e_{\gamma }\rfloor
^{(I-4)}Z^{\gamma }{}_{\beta })\right) \,.  \nonumber
\end{eqnarray}
Here $a_{0}$, $\ldots $, $a_{3}$, $b_{1}$, $\ldots $, $b_{5}$, $c_{2}$, $%
c_{3}$, $c_{4}$, $w_{1}$, $\ldots $, $w_{7}$, $z_{1}$, $\ldots $, $z_{9}$
are dimensionless coupling constants, $\kappa $ is the standard
gravitational constant, and $\rho $ is the strong gravity coupling constant.
We have introduced in the curvature square term the irreducible pieces of
the antisymmetric part $W_{\alpha \beta }\equiv R_{[\alpha \beta ]}$ and the
symmetric part $Z_{\alpha \beta }\equiv R_{(\alpha \beta )}$ of the
curvature two--form. In $Z_{\alpha \beta }$, we have the purely {\em post}%
--Riemannian part of the curvature. Note the peculiar cross terms with $%
c_{I} $ and $b_{5}$. The signature of space--time is $(-,+,+,+)$, the volume
four--form $\eta \equiv {}^{\ast }\!\,1$, and the two--form $\eta _{\alpha
\beta }\equiv \,^{\ast }(\vartheta _{\alpha }\wedge \vartheta _{\beta })$.

Therefore, space--time is described by a metric--affine geometry with the
gravitational {\em field strengths} nonmetricity $Q_{\alpha \beta }\equiv
-Dg_{\alpha \beta }$, torsion $T^{\alpha }\equiv D\vartheta ^{\alpha }$, and
curvature $R_{\alpha }{}^{\beta }\equiv d\Gamma _{\alpha }{}^{\beta }-\Gamma
_{\alpha }{}^{\gamma }\wedge \Gamma _{\gamma }{}^{\beta }$. The
gravitational field equations 
\begin{eqnarray}
DH_{\alpha }-E_{\alpha } &=&\Sigma _{\alpha }\,,  \label{first} \\
DH^{\alpha }{}_{\beta }-E^{\alpha }{}_{\beta } &=&\Delta ^{\alpha }{}_{\beta
}\,,  \label{second}
\end{eqnarray}
link the {\em material sources}, the material energy--momentum current $%
\Sigma _{\alpha }$ and the material hypermomentum current $\Delta ^{\alpha
}{}_{\beta }$, to the gauge field {\em excitations} $H_{\alpha }$ and $%
H^{\alpha }{}_{\beta }$ in a Yang--Mills--like manner. It is well known \cite
{PR} that the field equation corresponding to the variable $g_{\alpha \beta
} $ is redundant if (\ref{first}) as well as (\ref{second}) are fulfilled.
The excitations can be calculated by partial differentiation, 
\begin{equation}
H_{\alpha }=-\frac{\partial V_{{\rm MAG}}}{\partial T^{\alpha }}\,,\quad
H^{\alpha }{}_{\beta }=-\frac{\partial V_{{\rm MAG}}}{\partial R_{\alpha
}{}^{\beta }}\,,\quad M^{\alpha \beta }=-2\frac{\partial V_{{\rm MAG}}}{%
\partial Q_{\alpha \beta }}\,,\,  \label{3}
\end{equation}
whereas the gauge field currents of energy--momentum and hypermomentum,
respectively, turn out to be linear in the Lagrangian and in the
excitations, 
\begin{eqnarray}
E_{\alpha }\equiv &&\frac{\partial V_{{\rm MAG}}}{\partial \vartheta
^{\alpha }}=e_{\alpha }\rfloor V_{{\rm MAG}}+(e_{\alpha }\rfloor T^{\beta
})\wedge H_{\beta }+(e_{\alpha }\rfloor R_{\beta }{}^{\gamma })\wedge
H^{\beta }{}_{\gamma }+{\frac{1}{2}}(e_{\alpha }\rfloor Q_{\beta \gamma
})M^{\beta \gamma }\,, \\
E^{\alpha }{}_{\beta }\equiv &&\frac{\partial V_{{\rm MAG}}}{\partial \Gamma
_{\alpha }{}^{\beta }}=-\vartheta ^{\alpha }\wedge H_{\beta }-g_{\beta
\gamma }M^{\alpha \gamma }\,.
\end{eqnarray}
Here $e_{\alpha }$ represents the frame and $\rfloor $ the interior product
sign, for details see \cite{PR}. In vacuum, the energy--momentum current $%
\Sigma _{\alpha }=0$ and the material hypermomentum current $\Delta _{\beta
}^{\alpha }=0$.


\subsection{Triplet ansatz and equivalence theorem}

For the torsion and nonmetricity field configurations, we concentrate on the
simplest non--trivial case {\em with} shear. According to its irreducible
decomposition \cite{PR}, the nonmetricity contains two covector pieces,
namely $^{(4)}Q_{\alpha \beta }=Q\,g_{\alpha \beta }$, with $Q\equiv
g^{\alpha \beta }\,Q_{\alpha \beta }/4$, the dilation piece, and 
\begin{equation}
^{(3)}Q_{\alpha \beta }={\frac{4}{9}}\left( \vartheta _{(\alpha }e_{\beta
)}\rfloor \Lambda -{\frac{1}{4}}g_{\alpha \beta }\Lambda \right) \,,\quad 
\text{with}\quad \Lambda \equiv \vartheta ^{\alpha }e^{\beta }\rfloor \!{%
\nearrow \!\!\!\!\!\!\!Q}_{\alpha \beta }\,,  \label{3q}
\end{equation}
a proper shear piece. Accordingly, our ansatz for the nonmetricity reads 
\begin{equation}
Q_{\alpha \beta }=\,^{(3)}Q_{\alpha \beta }+\,^{(4)}Q_{\alpha \beta }\,.
\label{QQ}
\end{equation}
The torsion, in addition to its tensor piece, encompasses a covector and an
axial covector piece. Let us choose only the covector piece as
non--vanishing: 
\begin{equation}
T^{\alpha }={}^{(2)}T^{\alpha }={\frac{1}{3}}\,\vartheta ^{\alpha }\wedge
T\,,\quad \text{with}\quad T\equiv e_{\alpha }\rfloor T^{\alpha }\,.
\label{TT}
\end{equation}
Thus we are left with the three non--trivial one--forms $Q$, $\Lambda $, and 
$T$.

The Lagrangian (\ref{lagr}) is very complicated, in particular on account of
its curvature square pieces. Therefore we have to restrict its generality in
order to stay within manageable limits. Our ansatz for the nonmetricity is
expected to require a nonvanishing post--Riemannian term quadratic in the
segmental curvature. Accordingly, let be given the gauge Lagrangian (\ref
{lagr}) with $w_{1}=\dots =w_{7}=0$, $z_{1}=\dots =z_{3}=z_{5}=\dots
=z_{9}=0 $, that is, only $z_{4}$ is allowed to survive, i.e., the {\em %
segmental} curvature squared 
\begin{equation}
-\frac{z_{4}}{8\rho }\,R_{\alpha }{}^{\alpha }\wedge \,^{\ast }R_{\beta
}{}^{\beta }=-\frac{z_{4}}{2\rho }\,dQ\wedge {}^{\ast }dQ\,  \label{seg^2}
\end{equation}
is the only surviving strong gravity piece in $V_{{\rm MAG}}$.

We assume the following ansatz, the so--called {\em triplet ansatz} for our
triplet of one forms (\ref{QQ}) and (\ref{TT}): 
\begin{equation}
Q=k_{0}\,\phi \,,\qquad \Lambda =k_{1}\,\phi \,,\qquad T=k_{2}\,\phi \,,
\label{tripp}
\end{equation}
where $k_{0}\equiv 4\alpha _{2}\beta _{3}-3\gamma _{3}^{2}$, $k_{1}\equiv
9(\alpha _{2}\beta _{5}/2-\gamma _{3}\gamma _{4})$, $k_{2}\equiv 3(4\beta
_{3}\gamma _{4}-{3}\beta _{5}\gamma _{3}/2)$, and $\alpha _{2}=a_{2}-2a_{0}$%
, $\beta _{3}=b_{3}+a_{0}{/8}$, $\beta _{4}=b_{4}-3a_{0}{/8}$, $\beta
_{5}=b_{5}-a_{0}$, $\gamma _{3}=c_{3}+a_{0}$, $\gamma _{4}=c_{4}+a_{0}$. In
other words, we assume that the triplet of one--forms are proportional to
each other \cite{heh96,hema99,obu,tw,ghlms}.

The triplet ansatz (\ref{tripp}) reduces the MAG field equations (\ref{first}%
)--(\ref{second}) to an effective Einstein--Proca system \cite{obu,tw}: 
\begin{eqnarray}
\frac{a_{0}}{2}\,\eta _{\alpha \beta \gamma }\wedge \tilde{R}^{\beta \gamma
} &=&\kappa \,\Sigma _{\alpha }^{(\phi )},  \label{eq:Ein0} \\
d\,{}^{\ast }\!H+m^{2}\,{}^{\ast }\!\phi &=&0,  \label{eq:Proca0}
\end{eqnarray}
with respect to the metric $g$, and the Proca 1--form $\phi $. Here the
tilde\,$\tilde{\null}$\, denotes the Riemannian part of the curvature, 
\begin{eqnarray}
\Sigma _{\alpha }^{(\phi )} &\equiv &\frac{z_{4}k_{0}^{2}}{2\rho }\left\{
\left[ (e_{\alpha }\rfloor H)\wedge {}^{\ast }H\;-\;(e_{\alpha }\rfloor
\,^{\ast }H)\wedge {}H\right] \right.  \nonumber \\
&&\quad \left. +\;m^{2}\,\left[ (e_{\alpha }\rfloor \phi )\wedge {}^{\ast
}\phi \;+\;(e_{\alpha }\rfloor \,^{\ast }\phi )\wedge {}\phi \right]
\right\} \,,  \label{ProcaEM}
\end{eqnarray}
is the energy--momentum current of the Proca field $\phi $, $H\equiv d\phi $%
, and 
\begin{equation}
m^{2}\equiv \frac{1}{\kappa z_{4}}\left( -4\beta _{4}+\frac{k{_{1}}}{2k_{0}}%
\beta _{5}+\frac{k{_{2}}}{k_{0}}\gamma _{4}\right) ,
\end{equation}
Therefore, as mentioned above, given the triplet ansatz MAG becomes an
effective Einstein--Proca system. Moreover, by setting $m=0$ the system
acquires the constraint $\beta _{4}=(k{_{1}}\beta _{5}/2+k{_{2}}\gamma
_{4})/4k_{0}$ among the coupling constants of the Lagrangian (\ref{lagr}),
and it reduces to the Einstein--Maxwell system, cf. Ref. \cite{hema99}.

Thus, the triplet ansatz sector of a MAG theory becomes equivalent to the
Einstein--Proca system of differential equations. This was first shown for a
certain 3--parameter Lagrangian by Dereli et al. \cite{Dereli} and extended
to a 6--parameter Lagrangian by Tucker and Wang \cite{tw}. The situation was
eventually clarified for a fairly general 11--parameter Lagrangian by
Obukhov et al. \cite{obu}. In the next sections we will make use of this
equivalence theorem in order to proof the ``no--hair'' theorem for static
black--hole configurations in MAG.


\section{Black hole configurations in MAG}

The search for black hole solutions in MAG began with Tresguerres \cite
{TresguerresShear1,TresguerresShear2}, who found the first static
spherically symmetric solutions with a non--vanishing {\em shear charge},
i.e., the solutions are additionally endowed with a traceless part of the
nonmetricity. The metric of Tresguerres solution is the {\em %
Reissner--Nordstr\"om metric} of general relativity with cosmological
constant but the place of the electric charge is taken by the {\em dilation
charge} (gravito--electric charge) which is related to the trace of the
nonmetricity, the Weyl covector.

The Tresguerres solutions carry, besides the above--mentioned dilation and
shear charges (related to the trace and traceless pieces of the
nonmetricity, respectively), a {\em spin charge} related to the torsion of
space--time. Therefore, beyond the Reissner--Nordstr\"om metric, the
following post--Riemannian degrees of freedom are excited in the Tresguerres
solutions: {\em two} pieces of the nonmetricity, namely the Weyl covector $%
{}^{(4)}Q_{\alpha \beta }$ and the traceless piece ${}^{(2)}Q_{\alpha \beta
} $, and all {\em three} pieces of the torsion $^{(1)}T^{\alpha }$, $%
^{(2)}T^{\alpha }$, $^{(3)}T^{\alpha }$. The first solution \cite
{TresguerresShear1}, requires in the Lagrangian weak gravity terms and, for
strong gravity, the curvature square pieces with $z_{4}\neq 0$, $w_{3}\neq 0$%
, $w_{5}\neq 0$, i.e., with Weyl's segmental curvature, the curvature
pseudoscalar, and the antisymmetric Ricci.

Beside the two dilation--shear solutions, Tresguerres \cite
{TresguerresShear1} and Tucker and Wang \cite{tw} found
Reissner--Nordstr\"om--like metrics together with a non--vanishing Weyl
covector, $^{(4)}Q^{\alpha\beta}\neq 0$, and a vector part of the torsion, $%
^{(2)}T^\alpha\neq 0$, i.e., these solutions carry a {\em dilation} charge
(a Weyl charge) and a {\em spin} charge, but are devoid of any other
post--Riemannian ``excitations", in particular, they have no tracefree
pieces ${\nearrow\!\!\!\!\!\!\!Q}_{\alpha\beta}$ of the nonmetricity. The
corresponding Lagrangian needs only a Hilbert--Einstein piece $(a_0=1)$ and
a segmental curvature squared with $z_4 \neq 0$. The same has been proved
for the Tresguerres dilation solution \cite{TresguerresShear1} (see footnote
4 of \cite{heh96}).

In the framework of the triplet ansatz (\ref{tripp}), a
Reissner--Nordstr\"om--like metric with a strong gravito--electric charge
could successfully be used \cite{heh961} and a constraint on the coupling
constants had to be imposed. Thus the structure of this {\em triplet}
solution is reminiscent of the Tresguerres dilation--shear solutions.
Moreover, only the piece with $z_{4}\neq 0$ of the {\em curvature square
pieces} in the gauge Lagrangian $V_{{\rm MAG}}$ is required. All others do
not contribute. This result was generalized to an {\em axially symmetric}
solution \cite{heh961} based on the {\em Kerr--Newman} metric, with the same
kind of charges. A generalized Reissner--Nordstr\"{o}m solution in MAG
endowed with electric, magnetic, strong gravito--electric and strong
gravito--magnetic charges and cosmological constant is presented in \cite
{malo}.


\section{Uniqueness theorem for static black holes in metric--affine gravity}

For $m\neq 0$ the equivalence theorem \cite{obu,tw} establishes that the
triplet ansatz sector of MAG is described by the Einstein--Proca system (\ref
{eq:Ein0})--(\ref{eq:Proca0}). In this section we will proof that the
effective Proca field $\phi _\mu $ vanishes in the domain of outer
communications $\langle \!\langle 
\hbox{${\cal J}$\kern -.645em
{\raise.57ex\hbox{$\scriptscriptstyle (\ $}}}\rangle \!\rangle $ of a static
black hole. We would like to point out that the original proof on the
non--existence of massive--Proca fields in the presence of static black
holes is due to Bekenstein \cite{Bekens72b}. In this paper we improve the
original demonstration, basically in the arguments concerning the event
horizon, which are the more involved.

Using the standard component notation the Einstein--Proca equations (\ref
{eq:Ein0})--(\ref{eq:Proca0}) can be written as 
\begin{equation}
\frac{4\pi }{\widetilde{\kappa }}R_{\mu \nu }=H_{\mu }^{~\alpha }H_{\nu
\alpha }+m^{2}\phi _{\mu }\phi _{\nu }-\frac{1}{4}g_{\mu \nu }H_{\alpha
\beta }H^{\alpha \beta } \, ,  \label{eq:Ein}
\end{equation}
\begin{equation}
\nabla _{\beta }H^{\beta \alpha }=m^{2}\phi ^{\alpha },  \label{eq:Proca}
\end{equation}
where $R_{\mu \nu }$ is the Ricci tensor, $H_{\mu \nu }\equiv 2\nabla
_{\lbrack \mu }\phi _{\nu ]}$ is the field strength of the Proca field $\phi
_{\mu }$, and $\widetilde{\kappa }\equiv \kappa z_{4}k_{0}^{2}/\rho a_{0}$.

In a static black hole, the Killing field {\boldmath$k$} coincides with the
null generator of the event horizon ${{\cal {H}}}^{+}$ and is time--like and
hypersurface orthogonal in all the domain of outer communications $\langle
\!\langle 
\hbox{${\cal J}$\kern -.645em
{\raise.57ex\hbox{$\scriptscriptstyle (\ $}}}\rangle \!\rangle $. This allow
us to choose, by simply connectedness of $\langle \!\langle 
\hbox{${\cal J}$\kern -.645em
{\raise.57ex\hbox{$\scriptscriptstyle (\ $}}}\rangle \!\rangle $ \cite
{ChrWald95}, a global coordinates system $(t,x^{i})$,\, $i=1,2,3$, in all $%
\langle \!\langle 
\hbox{${\cal J}$\kern -.645em
{\raise.57ex\hbox{$\scriptscriptstyle (\ $}}}\rangle \!\rangle $ \cite
{Carter87}, such that $\text{\boldmath$k$}=\text{\boldmath$\partial
/\partial t$}$ and the metric reads 
\begin{equation}
\text{\boldmath$g$}=-V\text{\boldmath$dt$}^{2}+\gamma _{ij}\text{\boldmath$%
dx $}^{i}\text{\boldmath$dx$}^{j},  \label{eq:static}
\end{equation}
where $V$ and {\boldmath$\gamma $} are $t$--independent, {\boldmath$\gamma $}
is positive definite in all $\langle \!\langle 
\hbox{${\cal J}$\kern -.645em 
{\raise.57ex\hbox{$\scriptscriptstyle(\ $}}}\rangle \!\rangle $, and $V$ is
positive in all $\langle \!\langle 
\hbox{${\cal J}$\kern -.645em 
{\raise.57ex\hbox{$\scriptscriptstyle(\ $}}}\rangle \!\rangle $, and
vanishes in ${{\cal {H}}}^{+}$. From (\ref{eq:static}) it can be noted that
staticity is equivalent to the existence of a time--reversal isometry $%
t\mapsto -t$.

We will assume that the Proca field shares the same symmetries as the metric
one, namely, it is stationary, $\text{\boldmath${\pounds }_k\phi$}=0$. The
staticity of the metric is extended to the Proca field $\phi ^\alpha $ and
the Proca equations (\ref{eq:Proca}), i.e., they are invariant under
time--reversal transformations. The time--reversal invariance of Proca
equations (\ref{eq:Proca}) requires that, in the coordinates (\ref{eq:static}%
), $\phi ^t$ and $H^{ti}$ remain unchanged while $\phi ^i$ and $H^{ij}$
change their sign, or the opposite scheme, i.e., $\phi ^t$ and $H^{ti}$
change sign as long as $\phi ^i$ and $H^{ij}$ remain unchanged under time
reversal \cite{Bekens72b}. Therefore $\phi ^i$ and $H^{ij}$ must vanish in
the first case, and $\phi ^t$ and $H^{ti}$ vanish in the second one. Hence,
time--reversal invariance implies the existence of two separate cases: a
purely gravito--electric case (I) and a purely gravito--magnetic case (II).

Now we are ready to proof the ``{\em no--hair}'' theorem for the effective
Proca field. Let ${\cal V}\subset \langle \!\langle 
\hbox{${\cal J}$\kern -.645em
{\raise.57ex\hbox{$\scriptscriptstyle (\ $}}}\rangle \!\rangle $ be the open
region bounded by the space--like hypersurface $\Sigma $, the space--like
hypersurface $\Sigma ^{\prime }$ and the pertinent portions of the horizon 
{\em ${\cal H}^{+}$}, and the spatial infinity $i^o$. The space--like
hypersurface $\Sigma ^{\prime }$ is obtained by shifting each point of $%
\Sigma $ a unit parametric value along the integral curves of the Killing
field {\boldmath$k$}. Multiplying the Proca equations (\ref{eq:Proca}) by $%
\phi _\mu $ and integrating by parts over ${\cal {V}}$ using the Gauss
theorem, one obtains 
\begin{equation}
\left[ \int_{\Sigma ^{\prime }}-\int_\Sigma +\int_{{\cal {H}}^{+}\cap 
\overline{{\cal {V}}}}+\int_{i^o\cap \overline{{\cal {V}}}}\right] \phi
_\alpha H^{\beta \alpha }d\Sigma _\beta =\int_{{\cal {V}}}\left( \frac 12%
H_{\alpha \beta }H^{\alpha \beta }+m^2\phi _\alpha \phi ^\alpha \right) dv.
\label{eq:int}
\end{equation}
The boundary integral over $\Sigma ^{\prime }$ cancels out the corresponding
one over $\Sigma $, since $\Sigma ^{\prime }$ and $\Sigma $ are isometric
hypersurfaces. The boundary integral over $i^o\cap \overline{{\cal {V}}}$
vanishes by the usual Yukawa fall--off of the massive fields at infinity. We
will show that the integrand of the remaining boundary integral at the
portion of the horizon ${\cal {H}}^{+}\cap \overline{{\cal {V}}}$ also
vanishes. To achieve this goal we use the standard measure at the horizon $%
d\,\Sigma _\beta =2n_{[\beta }l_{\mu ]}l^\mu d\sigma $ \cite{Zannias98},
where {\boldmath$l$} is the null generator of the horizon, {\boldmath$n$} is
the other future--directed null vector ($n_\mu l^\mu =-1$), orthogonal to
the space--like cross sections of the horizon, and $d\sigma $ is the surface
element. The standard measure follows from choosing the natural volumen
3--form at the horizon, i.e., $\text{\boldmath$\eta _3$}={}^{*}(\text{%
\boldmath$n$}\wedge \text{\boldmath$l$})\wedge \text{\boldmath$l$}$. By
using the quoted measure the integrand over the horizon can be written as 
\begin{equation}
\phi _\alpha H^{\beta \alpha }d\Sigma _\beta =\left( \phi _\alpha H^{\beta
\alpha }l_\beta +\phi _\alpha H^{\beta \alpha }n_\beta \,l_\mu l^\mu \right)
d\sigma \,.  \label{eq:integ}
\end{equation}
In order to show the vanishing of the last integrand it is sufficient to
prove that the following quantities on the right hand side of (\ref{eq:integ}%
) are such that: $\phi _\alpha H^{\beta \alpha }l_\beta $ vanishes and $\phi
_\alpha H^{\beta \alpha }n_\beta $ remains bounded at the horizon. The
behavior of this quantities at the horizon can be established by studying
some invariants constructed from the curvature. By using Einstein equations (%
\ref{eq:Ein}), we obtain, 
\begin{equation}
\frac{4\pi }{\widetilde{\kappa }}R=m^2\phi _\mu \phi ^\mu ,
\label{eq:ScalarC}
\end{equation}
\begin{equation}
\frac{16\pi ^2}{\widetilde{\kappa }^2}R_{\mu \nu }R^{\mu \nu
}=3H^2+4I^2+\left( H-m^2\phi _\mu \phi ^\mu \right) ^2+2m^2H_\mu ^{~\alpha
}\phi ^\mu H_{\nu \alpha }\phi ^\nu ,  \label{eq:RicciS}
\end{equation}
where $H\equiv H_{\alpha \beta }H^{\alpha \beta }/4$, and $I\equiv
{}^{*}H_{\alpha \beta }H^{\alpha \beta }/4$. Since the horizon is a smooth
surface the left hand sides of the above equations are bounded on it. From (%
\ref{eq:ScalarC}) it follows that $\phi _\mu \phi ^\mu $ is bounded at the
horizon. The last term in (\ref{eq:RicciS}) is non--negative in both cases
(I) and (II), the remaining terms are also non--negative, and consequently
each one is bounded at the horizon, in particular the invariants $H$ and $I$%
. Other invariants can be built from the Ricci curvature (\ref{eq:Ein}) by
means of {\boldmath$l$} and{\ {\boldmath$n$}}, which are well--defined
smooth vector fields on the horizon. The first invariant reads 
\begin{equation}
\frac{4\pi }{\widetilde{\kappa }}R_{\mu \nu }n^\mu n^\nu =J_\mu J^\mu
+m^2(\phi _\mu n^\mu )^2-n_\mu n^\mu H\,,  \label{eq:Riccinn}
\end{equation}
where $J^\mu \equiv H^{\mu \nu }n_\nu $. The last term above vanishes
because the bounded behavior of the invariant $H$. Since {\boldmath$J$} is
orthogonal to the null vector {\boldmath$n$}, it must be space--like or null
($J_\mu J^\mu \geq 0$), therefore each one of the remaining terms on the
right hand side of (\ref{eq:Riccinn}) must be bounded. The next invariant to
be considered, which vanishes at the horizon by applying the Raychaudhuri
equation to the null generator \cite{Wald} reads 
\begin{equation}
0=\frac{4\pi }{\widetilde{\kappa }}R_{\mu \nu }l^\mu l^\nu =D_\mu D^\mu
+m^2(\phi _\mu l^\mu )^2-l_\mu l^\mu H\,,  \label{eq:Riccill}
\end{equation}
where $D^\mu \equiv H^{\mu \nu }l_\nu $ is the gravito--electric field at
the horizon. Once again the bounded behavior of the invariant $H$ can be
used to achive the vanishing of the last term of (\ref{eq:Riccill}). Since {%
\boldmath$D$} is orthogonal to the null generator {\boldmath$l$}, it must be
space--like or null ($D_\mu D^\mu \geq 0$), consequently each term on the
right hand side of (\ref{eq:Riccill}) vanishes independently, which implies
that $\phi _\mu l^\mu =0$ and that {\boldmath$D$} is proportional to the
null generator {\boldmath$l$} at the horizon, i.e., $\text{\boldmath$D$}%
=-(D_\alpha n^\alpha )\,\text{\boldmath$l$}$. The last studied invariant
gives the following relation, 
\begin{equation}
\frac{4\pi }{\widetilde{\kappa }}R_{\mu \nu }l^\mu n^\nu -H=(D_\mu n^\mu
)^2+m^2(\phi _\mu n^\mu )(\phi _\nu l^\nu ),  \label{eq:Ricciln}
\end{equation}
where it has been used that $\text{\boldmath$D$}=-(D_\alpha n^\alpha )\,%
\text{\boldmath$l$}$. Since $\phi _\mu l^\mu =0$ and $\phi _\mu n^\mu $ is
bounded at the horizon, it follows that the second term on the right hand
side of (\ref{eq:Ricciln}) vanishes, thus $D_\mu n^\mu $ is bounded at the
horizon as consequence of the bounded behavior of the left hand side of (\ref
{eq:Ricciln}).

Summarizing, the study of the quoted invariants at the horizon leads to the
following conclusions: $D_{\mu }n^{\mu }$, $\phi_{\mu }n^{\mu }$, $\phi
_{\mu }\phi ^{\mu }$, and $J_{\mu }J^{\mu }$ are bounded at the horizon, $%
\phi _{\mu }\,l^{\mu }=0$ and $\text{\boldmath$D$}=-(D_{\alpha }n^{\alpha
})\,\text{\boldmath$l$}$ in the same region.

Now we are in position to show the fulfillment of the sufficient conditions
for the vanishing of the integrand (\ref{eq:integ}) over the horizon, i.e., $%
\phi _{\alpha }H^{\beta \alpha }l_{\beta}$ vanishes and $\phi _{\alpha
}H^{\beta \alpha }n_{\beta}$ remains bounded at the horizon. Using the
definition $D^{\mu }\equiv H^{\mu \nu }l_{\nu }$ and that $\text{\boldmath$D$%
}=-(D_{\alpha }n^{\alpha })\,\text{\boldmath$l$}$, we obtain for the first
quantity at the horizon 
\begin{equation}
\phi _{\alpha }H^{\beta \alpha }l_{\beta }=(D_{\mu }n^{\mu })(\phi _{\nu
}l^{\nu })=0,  \label{eq:int1v}
\end{equation}
where the vanishing follows from the fact that $D_{\mu }n^{\mu }$ is bounded
and $\phi _{\nu }\,l^{\nu }$ vanishes at the horizon.

For the second quantity we note that {\boldmath$\phi $} and {\boldmath$J$}
are orthogonal to the null vectors {\boldmath$l$} and {\boldmath$n$},
respectively. Therefore, {\boldmath$\phi $} must be space--like or
proportional to {\boldmath$l$}, and {\boldmath$J$} must be space--like or
proportional to {\boldmath$n$}. Using a null tetrad basis, constructed with {%
\boldmath$l$}, {\boldmath$n$}, and a pair of linearly independent
space--like vectors, these last ones being tangent to the space--like cross
sections of the horizon, the {\boldmath$\phi $} and {\boldmath$J$} vectors
can be written as 
\begin{equation}
\text{\boldmath$\phi $}=-(\phi _\alpha n^\alpha )\text{\boldmath$l$}+\text{%
\boldmath$\phi $}^{\bot },  \label{eq:Bnt}
\end{equation}
\begin{equation}
\text{\boldmath$J$}=-(J_\alpha l^\alpha )\text{\boldmath$n$}+\text{\boldmath$%
J$}^{\bot },  \label{eq:Jnt}
\end{equation}
where $\text{\boldmath$\phi $}^{\bot }$ and $\text{\boldmath$J$}^{\bot }$
are the projections, orthogonal to {\boldmath$l$} and {\boldmath$n$}, on the
space--like cross sections of the horizon. Using (\ref{eq:Bnt}) and (\ref
{eq:Jnt}) it is clear that $\phi _\mu \phi ^\mu =\phi _\mu ^{\bot }\phi
^{\bot \mu }$, and $J_\mu J^\mu =J_\mu ^{\bot }J^{\bot \mu }$, i.e., the
contribution to these bounded magnitudes comes only from the space--like
sector orthogonal to {\boldmath$l$} and {\boldmath$n$}. With the help of (%
\ref{eq:Bnt}) and (\ref{eq:Jnt}) the other quantity appearing in the
integrand (\ref{eq:integ}) can be written as 
\begin{equation}
\phi _\alpha H^{\beta \alpha }n_\beta =-\phi _\alpha J^\alpha =-(\phi
_\alpha n^\alpha ){}(D_\beta n^\beta ){}-\phi _\alpha ^{\bot }J^{\bot \alpha
},  \label{eq:int2b}
\end{equation}
where the identity $J_\alpha l^\alpha =-D_\alpha n^\alpha $ has been used.
The first term in (\ref{eq:int2b}) is bounded because $\phi _\alpha n^\alpha 
$ and $D_\beta n^\beta $ are bounded. To the second term the Schwarz
inequality applies, since $\text{\boldmath$\phi $}^{\bot }$ and $\text{%
\boldmath$J$}^{\bot }$ belong to a space--like subspace. Thus, $(\phi
_\alpha ^{\bot }J^{\bot \alpha })^2\leq (\phi _\mu ^{\bot }\phi ^{\bot \mu
})(J_\nu ^{\bot }J^{\bot \nu })=(\phi _\mu \phi ^\mu )(J_\nu J^\nu )$ and
since $\phi _\mu \phi ^\mu $ and $J_\nu J^\nu $ are bounded at the horizon,
the second term of (\ref{eq:int2b}) is also bounded.

Finally, the vanishing of (\ref{eq:int1v}) and the bounded behavior of (\ref
{eq:int2b}), and the null character of {\boldmath$l$} at the horizon lead to
the vanishing of the integrand (\ref{eq:integ}) over the event horizon.

With no contribution from boundary integrals in (\ref{eq:int}) we shall
write the volume integral, using the coordinates from (\ref{eq:static}), for
each one of the different cases discussed at the beginning of this section.

For the purely gravito--electric case (I) we have 
\begin{equation}
\int_{{\cal {V}}}-V\left( \frac{1}{2}\gamma _{ij}H^{ti}H^{tj}+m^{2}(\phi
^{t})^{2}\right) dv=0.  \label{eq:zeroI}
\end{equation}
The non--positiveness of the above integrand, which is the sum of squared
terms, implies that the integral is vanishing only if $H^{ti}$ and $\phi
^{t} $ vanish everywhere in ${\cal {V}}$, and hence in all $\langle
\!\langle 
\hbox{${\cal J}$\kern -.645em
{\raise.57ex\hbox{$\scriptscriptstyle (\ $}}}\rangle \!\rangle $.

For the purely gravito--magnetic case (II) the volume integral reads as 
\begin{equation}
\int_{{\cal {V}}}\left( \frac{1}{2}\gamma _{ik}\gamma
_{jl}H^{kl}H^{ij}+m^{2}\phi _{i}\phi ^{i}\right) dv=0,  \label{eq:zeroII}
\end{equation}
in this case the non--negativeness of the above integrand is responsible for
the vanishing of $H^{ij}$ and $\phi ^{i}$ in all $\langle \!\langle 
\hbox{${\cal J}$\kern -.645em
{\raise.57ex\hbox{$\scriptscriptstyle (\ $}}}\rangle \!\rangle $.


\section{Discussion}

By using the equivalence theorem for the triplet sector of MAG, one faces in
vacuum an effective Einstein--Proca system, where the Proca field is related
to the vector pieces of the irreducible decomposition of the nonmetricity
and torsion \cite{obu,tw}. In this framework, we investigate whether or not
this sector of MAG leads to new physics, i.e., we analyze the ``{\em no--hair%
}'' problem in the triplet ansatz sector of MAG.

We prove that, for $(-4\beta _4+k_1\beta _5/2k_0+k_2\gamma _4/k_0)/\kappa
z_4\neq 0$, the effective field $\phi _\mu $ is trivial in the presence of a
static black hole. This result implies that the equations (\ref{eq:Ein0})--(%
\ref{eq:Proca0}) reduce to the Einstein--vacuum ones, for which the only
static black hole is the Schwarzschild solution \cite{Israel67}. For the
special case where $(-4\beta _4+k_1\beta _5/2k_0+k_2\gamma _4/k_0)/\kappa
z_4=0$, the system (\ref{eq:Ein0})--(\ref{eq:Proca0}) reduces to the
Einstein--Maxwell system, and it is well--known that the only static black
hole, with non--degenerate horizon, is the Reissner--Nordstr\"om one \cite
{Israel68} (see Refs. \cite{Heusler96,Heusler98} for improvements to the
original proofs). It is worthwhile to stress the fact that while previous
works by Chru\'sciel and Nadirashvili \cite{ChruscielN95}, and by Heusler
\cite{Heusler97}, on the establishment of the uniqueness results in the 
extremal case were not yet conclusive, the problem seems to
be settled recently by Chru\'sciel \cite{Chrusciel99}, who establishes the
uniqueness of the Majundar--Papapetrou black hole in the extremal case.

It is straightforward to generalize our results to static electrovacuum
space--times of the triplet sector of MAG, which in view of the equivalence
theorem becomes equivalent to an effective Einstein--Proca--Maxwell system,
i.e., the only existing static black hole is the true Reissner--Nordstr\"om
one, endowed with electric and/or magnetic charge as well as
gravito--electric and/or gravito--magnetic charges \cite{malo}. The proof
involves only and additional Maxwell field.

Moreover, the existence of static soliton (particle--like) solutions can be
also excluded using the same arguments, since the only change in the proof
is that in this case the boundary of the volume ${\cal {V}}$ is only formed
by the isometric surfaces $\Sigma $ and $\Sigma ^{\prime }$, and a portion
of the spatial infinity $i^o$, i.e., there are no interior boundary
corresponding to the event horizon.


\acknowledgments
We thank Friedrich W. Hehl, for useful discussions and literature hints.
This research was partially supported by CONACyT Grants: ``Black Holes,
Inflation and Matter'', 28339E, 3567E, DGAPA--UNAM Grant 121298, and by a
fellowship from the Sistema Nacional de Investigadores (SNI).


\end{document}